\documentclass[aip,jmp]{revtex4-1}

\usepackage{amsmath}
\usepackage{amssymb}
\usepackage{graphicx}
\usepackage{color}
\usepackage{textcomp} 
\usepackage{bm} 

\usepackage[implicit=true,colorlinks=true,linkcolor=blue,citecolor=blue,urlcolor=blue]{hyperref}

\newcommand{\fracd}[3][...]{\displaystyle\frac{\mathrm{d}^{#1}#2}{\mathrm{d}#3^{#1}}}
\newcommand{\fracp}[3][...]{\displaystyle\frac{\partial^{#1}#2}{\partial#3^{#1}}}

\newcommand{\dd}{\mathrm{d}}
\newcommand{\ii}{\mathrm{i}}
\newcommand{\ee}{\mathrm{e}}

\newcommand{\SE}{Schr\"odinger equation}
\newcommand{\up}{\mathrm{u}}
\newcommand{\lw}{\mathrm{l}}

\usepackage[normalem]{ulem} 
\definecolor{uniorange}{RGB}{250, 100, 55}
\definecolor{darkgreen}{rgb}{0.2,.7,0.2}

\begin{document}

\title{Recursive formulation of Madelung continuity equation leads to propagation equation}

\author{D.\,M.~Heim}
\email{physics@d-heim.de}
\affiliation{Fraunhofer Institute for Industrial Mathematics, 67663 Kaiserslautern, Germany}
\affiliation{Key Lab of Quantum Information of Yunnan Province, Yunnan University, Kunming, Yunnan, 650091 China}

\date{\today}

\begin{abstract}
We apply a recursive approach to the continuity equation of the Madelung fluid  resulting in a propagation equation for particle probability densities.
This propagation equation can be used to propagate particle distributions in the presence of a Madelung pressure field.
We show that the derived propagation equation goes over into the guidance equation of the de Broglie-Bohm theory in the limit of well located single particles.
As an example, we propagate particles that enter the lower slit of a double-slit experiment, while the Madelung fluid enters both slits.
\end{abstract}



\maketitle

\section{Introduction}\label{sec:introduction}

The concept of a hydrodynamical model of quantum mechanics was first proposed in 1926 by Madelung \cite{madelung:1926}, followed by the proposal of Korn in 1927 \cite{korn:1927}.
This concept revealed to be very fruitful in a number of applications and is now established in several branches of quantum mechanics \cite{wyatt:2005}, such as Bose-Einstein-condensation \cite{pitaevskii:2016,engels:2017}, condensed matter physics \cite{fedele:2006,wei:2007} and quantum cosmology \cite{pashaev:2010,carusotto:2016}.
It has been also applied as a useful tool to solve linear and non-linear partial differential equations \cite{nonnenmacher:1987,fedele:2009,malomed:2018,koeppe:2018}. We extend the Madelung fluid description by applying a recursive approach to its continuity equation which leads to a propagation equation.

The Madelung equations arise when the quantum mechanical wave function in polar decomposition
\begin{align}
			\psi \equiv \sqrt{P}\ \ee^{{\ii}{} S }
			\label{eq:wave-function-polar}
\end{align}
is inserted into the \SE{}
\begin{align}
	\ii \hbar \fracp[]{}{t} \psi = \left( -\frac{\hbar^2}{2m} \mathbf{\nabla}^2 + V \right) \psi
	\label{eq:schroedinger:full}
\end{align}
and the imaginary part
\begin{align}
	\fracp[]{}{t} {P} &= - \frac{\hbar}{m} \mathbf{\nabla} ( P\, \nabla S )
	\label{eq:imag:s}
\end{align}
and the real part
\begin{align}
	\hbar \fracp[]{}{t} {S} = \frac{\hbar^2}{2m} 
	\left[ \frac{\nabla^2 \sqrt{P}}{\sqrt{P}} - (\nabla S)^2\right] - V
	\label{eq:real-part}
\end{align}
are separated. Here, we used the probability density $P = P(\mathbf{r},t)$ to find a particle of mass $m$ at the position $\mathbf{r}$ and time $t$, and the phase $S=S(\mathbf{r},t)$.

By defining a velocity field
\begin{align}
	\mathbf{v}=\mathbf{v}(\mathbf{r},t) \equiv \frac{\hbar}{m} \nabla S
	, \label{eq:velocity-field}
\end{align}
the imaginary part \eqref{eq:imag:s} can be written in the form of a continuity equation
\begin{align}
	\fracp[]{}{t} {P} &= - \mathbf{\nabla} ( \mathbf{v}\, P )
	\label{eq:imag}
\end{align}
while the gradient of the real part \eqref{eq:real-part}
\begin{align}
	\fracp[]{}{t} {\mathbf{v}} + \mathbf{v} \nabla \mathbf{v} = - \frac{1}{m} \nabla \left( V + Q  \right)
	\label{eq:euler}
\end{align}
is similar to the Euler equation of classical hydrodynamics \cite{landau:1987}. Here, we defined the quantum pressure field
\begin{align}
	Q \equiv - \frac{\hbar^2}{2m} 
	\frac{\nabla^2 \sqrt{P}}{\sqrt{P}}
	. \label{eq:presssure}
\end{align}
Consequently, we obtain a set of equations which describe the motion of an inviscid fluid and can be used to predict the outcome of non-relativistic quantum experiments \cite{wyatt:2005}.

We extend the Madelung fluid description by re-writing its continuity equation \eqref{eq:imag} as a propagation equation
	\begin{align}
		P = \widehat{G}[\mathbf{v}]\, P_0
		, \label{eq:propagation:propagator}
	\end{align}
which only involves an initial probability density {$P_0 \equiv P(\mathbf{r},t_0)$} and the velocity field \eqref{eq:velocity-field} as argument of the propagator $\widehat{G}$, which we derive in this publication.

The propagation equation \eqref{eq:propagation:propagator} is useful to describe the motion of particles in the presence of a Madelung pressure field.
We show that this equation interestingly goes over into the guidance equation of the de Broglie-Bohm theory \cite{bohm:1952:1} for single particles.
As an example, we apply the propagation equation in the case of a double-slit experiment{, which illustrates the connections of our formalism with quantum mechanics and the de Broglie-Bohm theory}.

\section{Propagation equation} \label{sec:propagation-equation}

In order to derive the propagation equation \eqref{eq:propagation:propagator}, we integrate the continuity equation \eqref{eq:imag}, which leads us with the abbreviations $\mathbf{v}_n \equiv \mathbf{v}(\mathbf{r},t_n)$ and $P_n \equiv P(\mathbf{r},t_n)$ to
\begin{align}
P = P_0 \,- \int\limits_{t_0}^{t}\dd t_1\ \mathbf{\nabla} ( {\mathbf{v}_1}\, P_1 ) .\label{eq:imaginary:first}
\end{align}

Similar to deriving a formal solution of the \SE{} \cite{abrikosov1964}, we reinsert equation~\eqref{eq:imaginary:first} multiple times into the right-hand-side of itself which leads to the series
\begin{align}
{P} &= {P}_0 {\,-} \int\limits_{t_0}^{t}\dd t_1\ \mathbf{\nabla} \left( {{\mathbf{v}_1}}\,{P}_0  \right) 
+ \int\limits_{t_0}^{t}\dd t_1\ \mathbf{\nabla} {{\mathbf{v}_1}}
\int\limits_{t_0}^{t_1}\dd t_2\ \mathbf{\nabla} \left( {{\mathbf{v}_2}}\,{P}_0  \right) - \ldots
\label{eq:series}
\end{align}
and finally to the propagation equation \eqref{eq:propagation:propagator}, that is
\begin{align}
	P = \widehat{G}[\mathbf{v}]\, P_0
	, \label{eq:propagation:time-sorting}
\end{align}
where we defined the propagator
\begin{align}
	\widehat{G}[\mathbf{v}]  \equiv \hat{T} \left[ \exp \left( {-} \int_{t_0}^{t}\dd t_1\ \mathbf{\nabla} {{\mathbf{v}_1}} \right) \right]
	. \label{eq:propagator:definition}
\end{align}
The operator $\hat{T}$ is the well-known time-ordering operator, which is defined by
\begin{align}
\hat{T} \left[ \hat{A}(t)\,\hat{B}(t') \right] \equiv
\begin{cases}
\hat{A}(t)\,\hat{B}(t')\qquad \text{for } \,t\,\geq t' \\
\hat{B}(t')\,\hat{A}(t)\qquad \text{for } t'\leq t
\end{cases}
. \label{eq:time-ordering}
\end{align}

Equation \eqref{eq:propagation:time-sorting} describes the propagation of an initial probability density ${P}_0$ to the final density ${P}$ by using the velocity field $\mathbf{v}$.

\section{Propagation of single particle} \label{sec:bohm}

In this section, we show that the propagation equation \eqref{eq:propagation:time-sorting} goes over into the guidance equation of the de Broglie-Bohm theory in the limit of well located single particles.

{
The de Broglie-Bohm theory was originally proposed by de Broglie in 1927 \cite{broglie:1927} and refined by Bohm in 1952 \cite{bohm:1952:1,bohm:1952:2}. This theory supplements quantum mechanics by classical particle trajectories. This is done by regarding the real part \eqref{eq:real-part} of the \SE{} as Hamilton-Jacobi equation
\begin{align}
	\fracp[]{}{t} {\hbar S} + \frac{(\nabla \hbar S)^2}{2 m} + V + Q = 0
	, \label{eq:hamilton-jacobi}
\end{align}
where $Q$ is called the quantum potential and $\hbar S$ the Hamilton-Jacobi function. The trajectory $\bm{\xi}(t)$ of a single particle is then defined in analogy to the Hamilton-Jacobi formalism by solving the first-order differential equation
\begin{align}
	\fracd[]{\bm{\xi}(t)}{t} = \frac{1}{m} \nabla \hbar S
	, \label{eq:guidance-equation}
\end{align}
which is known as the guidance equation\cite{holland:1995}.
}

{To reproduce this equation from our formalism}, we replace the probability density $P$ in the propagation equation~\eqref{eq:propagation:time-sorting} by the Dirac delta function
\begin{align}
 P \rightarrow \delta[\mathbf{r}-\bm{\xi}(t)]
 ,
\end{align}
where $\bm{\xi}(t)$ represents the trajectory of a single particle. After multiplying the resulting equation by $\mathbf{r}$ and integrating over all space we obtain
\begin{align}
	\bm{\xi}(t) = \int \dd V\ \mathbf{r}\ \widehat{G}[\mathbf{v}]\, \delta[\mathbf{r}-\bm{\xi}(t_0)]
	. \label{eq:propagation:dirac:multiplied}
\end{align}
To rewrite the probability density $\widehat{G}[\mathbf{v}]\, \delta[\mathbf{r}-\bm{\xi}(t_0)]$ we perform the steps from equation~\eqref{eq:propagation:time-sorting} to equation~\eqref{eq:imaginary:first} in reverse which leads us to
\begin{align}
	\bm{\xi}(t) = \bm{\xi}(t_0) \,- \int\limits_{t_0}^{t}\dd t_1\ \int \dd V\ \mathbf{r}\ \mathbf{\nabla} \left\{ {\mathbf{v}_1}\, \delta[\mathbf{r}-\bm{\xi}(t_1)] \right\}
	.\label{eq:imaginary:dirac}
\end{align}
Here, we perform a partial integration to obtain
\begin{align}
 \bm{\xi}(t) = \bm{\xi}(t_0) + \int\limits_{t_0}^{t}\dd t_1\, \mathbf{v}[\bm{\xi}(t_1),t_1]
 . \label{eq:pre-guidance}
\end{align}
{By inserting the velocity field \eqref{eq:velocity-field} into \eqref{eq:pre-guidance} and taking the derivative with respect to $t$ we arrive at the guidance equation \eqref{eq:guidance-equation}}
of the de Broglie-Bohm theory.
{We compare the application of the propagation equation \eqref{eq:propagation:time-sorting} with \eqref{eq:pre-guidance} in the next section.}

\section{Propagation of particle distribution} \label{sec:double-slit}

\begin{figure}[t]
\begin{center}
	\includegraphics[]{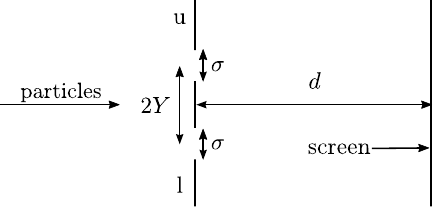}
\end{center}
	\caption{Double-slit experiment, where particles can pass two slits denoted by u and l with a distance $2Y$. Each slit has a width $\sigma$ and distance $d$ from the screen where the particles impinge. }
	\label{fig:setup}
\end{figure}

In the following, we propagate particle probability densities in a double-slit experiment which we sketched in Fig.~\ref{fig:setup}. We {compare the quantum mechanical probability density with trajectories arising from the Broglie-Bohm theory and with a probability density resulting from a thought experiment, where we} consider a Madelung fluid that enters both slits while the particles pass only the lower slit.

{
\subsection{Quantum mechanics}
\label{sec:quantum}
}

The wave function
\begin{align}
  {{\psi}}(y,t) = \frac{{\psi}_{\up}(y,t) + {\psi}_{\lw}(y,t)}{\sqrt{2}}
  , \label{eq:final-fuction-time}
\end{align}
which describes the possibility {of finding particles behind the slits in Fig.~\ref{fig:setup},} consists of the sum of the wave functions ${\psi}_{\up}(y,t)$ passing the upper and ${\psi}_{\lw}(y,t)$ passing the lower slit.

We represent these wave functions by Gaussians
\begin{align}
	{\psi}_{\up,\lw}(y,t) =  {\frac{1}{\sqrt[4]{2\pi}\sqrt{\sigma+\ii {\hbar t}/{({2}m{\sigma}})}}} 
	 \exp  \left[ -\frac{(y\mp Y )^2}{4{{\sigma}_{}}^2+2\ii {\hbar t}/m} \right] 
	 \label{eq:final-functions-single-time}
\end{align}
which solve the \SE{} \eqref{eq:schroedinger:full} for $V=0$. Here, $2Y$ is the slit separation and $\sigma$ describes how much the probability density expands in a single slit.

\begin{figure}[t]
	\begin{center}
		\includegraphics[]{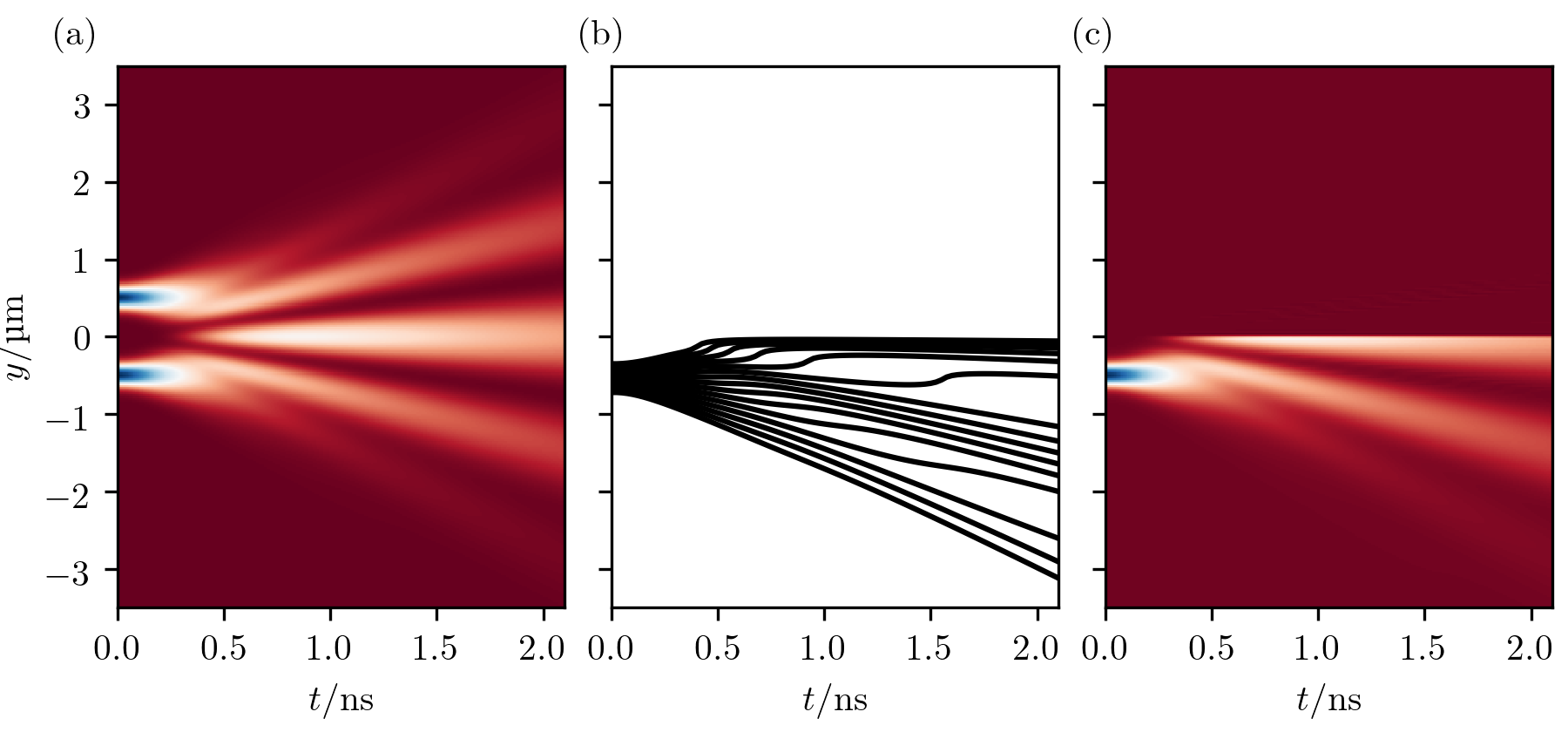}
	\end{center}
	\caption{
		{Application of different formalisms in the double-slit setup shown in Fig.~\ref{fig:setup}. In~(a) we depicted the quantum mechanical probability density $|{\psi}(y,t)|^2$ of the wave function \eqref{eq:final-fuction-time}, (b) shows de Broglie-Bohm trajectories calculated from equation \eqref{eq:guidance} and the probability density in~(c) is calculated by using the propagation equation \eqref{eq:propagation:set}. From (c) one can see that this propagation equation can describe continuous probability densities as in quantum mechanics (a), but also offers the possibility to propagate only parts of the initial probability density as it can be done in the de Broglie-Bohm theory (b).}
		}
	\label{fig:interference}
\end{figure}

{
In Fig.~\ref{fig:interference}(a), we used the wave function \eqref{eq:final-fuction-time} to depict the quantum mechanical probability density $|{\psi}(y,t)|^2$ for parameters similar to the ones of the experiment performed by J\"onsson et al. \cite{joensson:1961}. The slit separation is taken to be $2Y=1$\,\textmu m and $\sigma=0.1$\,\textmu m. After a short propagation time $t$, the probability density displays the well known interference fringes.
}

{
\subsection{De Broglie-Bohm}
\label{sec:bohm-example}
}

{
To calculate de Broglie-Bohm trajectories, we select 30 positions $y_j(0)$ for the time $t_0=0$ at the location of the lower slit in Fig.~\ref{fig:setup}. We propagate each of these positions by using equation \eqref{eq:pre-guidance} in the form
\begin{align}
 y_j(t) = y_j(0) + \int\limits_{0}^{t}\dd t_1\, \mathbf{v}[y(t_1),t_1]
 .\label{eq:guidance}
\end{align}
Here, we used the wave function \eqref{eq:final-fuction-time} in the velocity field \eqref{eq:velocity-field} which leads us to
\begin{align}
\mathbf{v}(y,t) = \frac{\hbar}{m} \fracp[]{}{y} \mathrm{arg}[{{\psi}}(y,t)]
. \label{eq:velocity:double}
\end{align}
The resulting trajectories are shown in Fig.~\ref{fig:interference}(b) for the same parameters as we used in Fig.~\ref{fig:interference}(a).
}

{As already reported by Bohm and Hiley\cite{bohm:1975}, the trajectories in such a double-slit experiment are arranged along the valleys of the interference pattern.
Differently from their publication, we selected only trajectories emerging from the lower slit in order to stress the similarity between our formalism and the de Broglie-Bohm theory which we discuss in more detail in the next section. However, to reproduce experimental results, the distribution of initial positions should match the density $|{\psi}(y,0)|^2$ which covers both slits.
}

{
\subsection{Madelung fluid}
}

{For the calculation of the probability density resulting from a Madelung pressure field \eqref{eq:presssure}, we use the wave function ${\psi}_{\lw}(y,t)$ defined in \eqref{eq:final-functions-single-time} to describe an initial probability density
\begin{align}
 {\rho}_0 \equiv |{\psi}_{\lw}(y,0)|^2
  \label{eq:probabiliy:lower}
\end{align}
located at the lower slit in Fig.~\ref{fig:setup}. With the help of the propagation equation \eqref{eq:propagation:time-sorting}, we determine its time evolution
\begin{align}
	{\rho} = \widehat{G}[\mathbf{v}]\, {\rho}_0
	, \label{eq:propagation:set}
\end{align}
}
{where we use the the wave function \eqref{eq:final-fuction-time} in the velocity field \eqref{eq:velocity:double} to describe the Madelung fluid.
}

{
We depicted the resulting probability density ${\rho}$ in Fig.~\ref{fig:interference}(c) for the same parameters as we used in Figs.~\ref{fig:interference}(a) and~\ref{fig:interference}(b).}
{With increasing time}, the probability density ${\rho}$ still covers only values in the lower part behind the double-slit, but interference can be observed. Similar patterns were created by Zheng et al.~\cite{zheng:2012} {with the help of variant simulations}. 
In Fig.~\ref{fig:interference}(c) it looks as if the upper maximum is not connected to the others, but there is just a low probability to find the particles. In a de Broglie-Bohm interpretation of quantum mechanics this would mean that the particles move fast at these positions.

{
To reproduce results of physical double-slit experiments, the initial probability density $\rho_0$ must match the squared modulus $|\psi(y,0)|^2$ of the quantum mechanical wave function which is used in the velocity field \eqref{eq:velocity:double}.
This is not fulfilled by the initial probability density \eqref{eq:probabiliy:lower} of our thought experiment. Nevertheless, this thought experiment is very interesting from a mathematical point of view.
}

{It shows that our formalism combines characteristics of quantum mechanics with characteristics of the de Broglie-Bohm theory. This can be seen from Fig.~\ref{fig:interference}, where the result of our formalism~(c) describes continuous probability densities as in quantum mechanics~(a), but also offers the possibility to propagate parts of the initial probability density as in the de Broglie-Bohm theory~(b).
}

\hspace{2cm}

\section{Summary and conclusion}\label{sec:discussion}

In summary, we have rewritten the Madelung continuity equation as a propagation equation which can be used to propagate particle probability densities in a Madelung fluid. We have shown that the derived propagation equation goes over into the guidance equation of the de Broglie-Bohm theory, where only single particles are propagated. {As an example, we illustrated similarities of our formalism with quantum mechanics and with the de Broglie-Bohm theory by propagating particle probability densities in a double-slit experiment.}

A possible application of our formalism is to simplify the calculation of multiple trajectories in the de Broglie-Bohm theory. Here, instead of calculating each trajectory individually, the derived propagation equation can be used to propagate a distribution of multiple trajectories at once.

Since our approach extends the use of the Madelung equations, it can also be a helpful tool for its fields of application, e.g. the description of Bose-Einstein condensates, the physics of condensed matter and quantum cosmology. In addition, our recursive approach to the continuity equation can be helpful not only for probabilities, but also for systems that describe mass, energy, electric charge or heat flows.

\section{Acknowledgment} \label{sec:acknowledgement}

I thank Prof.~J.~Zheng for initiating my interest in formulations of quantum mechanics and for many stimulating discussions on this topic. Furthermore, I thank Prof. W.~P.~Schleich, S.~V.~Aksenov, J.~Fischbach, C.~K.~Tempel and R.~Heese for critical and very fruitful discussions and O.~Heim for revising the manuscript. Financial support by the National Natural Science Foundation of China NSFC (No. 61362014) and the Overseas Higher-level Scholar Project of Yunnan Province, China, is gratefully acknowledged.

\end{document}